\documentclass[12pt]{article}
\usepackage{graphicx}
\usepackage{amssymb}
\usepackage{amsmath}

\newcommand{\bear}{\begin{array}}  
\newcommand {\eear}{\end{array}}
\newcommand{\bea}{\begin{eqnarray}}   
\newcommand{\eea}{\end{eqnarray}}
\newcommand{\beq}{\begin{equation}}   
\newcommand{\eeq}{\end{equation}}
\newcommand{\bef}{\begin{figure}}  \newcommand 
{\eef}{\end{figure}}
\newcommand{\bec}{\begin{center}}  \newcommand 
{\eec}{\end{center}}
  
\newcommand{\la}{\left\langle}  
\newcommand{\ra}{\right\rangle}

\def\GEV#1{10^{#1}{\rm\,GeV}}

\begin{document}

\begin{titlepage}

\begin{flushright}
KEK-TH-1382\\
IPMU10-0127\\
\end{flushright}

\vskip 1.35cm


\begin{center}

{\large \bf
Constraint on the gravitino mass in hybrid inflation
}

\vskip 1.2cm

Kazunori Nakayama$^a$,
Fuminobu Takahashi$^b$
and 
Tsutomu T. Yanagida$^{b,c}$

\vskip 0.4cm

{ \it $^a$Theory Center, KEK, 1-1 Oho, Tsukuba, Ibaraki 305-0801, Japan}\\
{\it $^b$Institute for the Physics and Mathematics of the Universe,
University of Tokyo, Kashiwa 277-8568, Japan}\\
{\it $^c$Department of Physics, University of Tokyo, Tokyo 113-0033, Japan}

\date{\today}

\begin{abstract} 
We revisit the $F$-term hybrid inflation model in supergravity.
In particular, we point out that a constant term in the superpotential has 
significant effects on the inflaton dynamics.
It is shown that the hybrid inflation model suffers from 
several potential problems: tuning of the initial condition, gravitino overproduction and
formation of cosmic strings, for both minimal and non-minimal K\"ahler potentials.
These problems can only be avoided in gauge-mediated SUSY breaking models
where the gravitino is relatively light and the constant term in the superpotential is not important.
Implications on the non-thermal leptogenesis scenario are also described.
\end{abstract}



\end{center}
\end{titlepage}

\section{Introduction}

The existence of inflationary expansion era in the early Universe is strongly motivated
by cosmological observations~\cite{Komatsu:2010fb}.
On the other hand, supersymmetry (SUSY) is the most plausible candidate for new physics
from naturalness point of view.
Various inflation models based on SUSY or supergravity setup have been proposed so far,
including hybrid inflation~\cite{Copeland:1994vg,Dvali:1994ms,Binetruy:1996xj,Linde:1997sj}, 
new inflation~\cite{Izawa:1996dv} 
and chaotic inflation~\cite{Kawasaki:2000yn,Kadota:2007nc,Takahashi:2010ky}.

Another hint for the new physics beyond the standard model may be the smallness of the neutrino masses as measured by neutrino oscillation experiments and constrained cosmologically.
The tiny neutrino masses are generated by integrating out heavy right-handed neutrinos
through the see-saw mechanism~\cite{Yanagida:1979}.
In order to give heavy masses to right-handed neutrinos, we introduce the following superpotential
\begin{equation}
	W = y_{ij} \frac{1}{2}\Psi N_i N_j,   \label{PNN}
\end{equation}
where $N_i$ is the right-handed neutrino superfield with $i$ 
denoting the generation number $(i=1,2,3)$,
$\Psi$ is a singlet under the standard model gauge group and $y_{ij}$ is a coupling constant.
This form of the potential is ensured by the gauged $U(1)_{B-L}$ symmetry under which
$\Psi$ and $N_i$ have charges of $+2$ and $-1$.\footnote{
	Instead, if one assigns a $U(1)_{B-L}$ charge of $+1$ for $\Psi$, the non-renomalizable superpotential
	$W \sim \Psi\Psi N_iN_j / \Lambda$ with cutoff scale $\Lambda$ is allowed.
}
If the scalar component of $\Psi$ gets a large vacuum expectation value (VEV),
the right-handed (s)neutrinos obtain large masses, leading to 
tiny left-handed neutrino masses through the see-saw mechanism.
The simplest superpotential for giving $\Psi$ a large VEV is the following,
\begin{equation}
	W = \kappa \Phi(\Psi \bar \Psi - M^2), \label{kSPP}
\end{equation}
where $\kappa$ is a coupling constant, $\Phi$ is a singlet field,
and $\bar \Psi$ has a charge $-2$ under $U(1)_{B-L}$.
Along with the $D$-term potential which enforces $|\Psi| = |\bar \Psi| $,
the vacuum lies at $\Phi =0$ and $|\Psi| = |\bar \Psi| = M$, and hence right-handed (s)neutrinos
obtain large masses.
Remarkably, the superpotential (\ref{kSPP}) exactly corresponds to
that for hybrid inflation~\cite{Dvali:1994ms},
with $\Phi$ identified as the inflaton and $\Psi (\bar \Psi)$ as the waterfall field.

Thus we are led to consider a SUSY hybrid inflation model.
The hybrid inflation model in supergravity was studied in detail in Refs.~\cite{Lazarides:1996dv,Buchmuller:2000zm,Senoguz:2004vu,Jeannerot:2005mc,BasteroGil:2006cm,urRehman:2006hu,Pallis:2009pq}.
In this paper we reexamine the hybrid inflation model in various aspects
and explore the parameter space where various constraints are satisfied, and 
discuss issues concerning fine-tuning of the initial condition for successful inflation.
In particular, we focus on the effect of a constant term in the superpotential
on the inflaton dynamics,
which is required for making the cosmological constant zero.
While earlier works~\cite{Buchmuller:2000zm,Senoguz:2004vu} focused on the limited parameter range,
we consider both minimal and non-minimal K\"ahler potentials with wide range of the gravitino mass.
We also explicitly show which parameter regions suffer from the fine-tuning of the initial condition.
Moreover, we take into account the non-thermal gravitino production from inflaton 
decay~\cite{Kawasaki:2006gs,Endo:2007ih,Asaka:2006bv,Endo:2006tf},
which places an extremely tight constraint on the inflation model parameters for an unstable TeV-scale gravitino,
allowing only the gravitino mass either heavier than $O(10)$TeV or lighter than $O(1)$GeV.
Interestingly, as demonstrated in the following sections, the inclusion of the constant term in the superpotential
gives tight and complementary constraints for the heavier gravitino, which, together with the gravitino constraint,  gives a clear preference to a light gravitino.

This paper is organized as follows.
In Sec.~\ref{sec:model} we derive the potential of the hybrid inflation model and
show what kinds of constraints should be considered.
In Sec.~\ref{sec:const} we present the resultant constraints on the parameter space.
In Sec.~\ref{sec:lepton} implications on non-thermal leptogenesis scenario are described.
Sec.~\ref{sec:conc} is devoted to conclusions.

\section{$F$-term hybrid inflation model}   \label{sec:model}

\subsection{Scalar potential}
\label{scalar potential}

The model consists of the singlet superfield $\Phi$ and
U(1)$_{B-L}$ Higgs fields $\Psi (\bar \Psi)$ with charge $+2 (-2)$.
The $F$-term hybrid inflation is based on the following superpotential and K\"ahler potential,
\begin{gather}
	K =  |\Phi|^2 + |\Psi|^2 +|\bar \Psi|^2 + k_1 \frac{|\Phi|^4}{M_P^2} 
		+k_2\frac{|\Phi|^2 |\Psi|^2}{M_P^2}+k_3\frac{|\Phi|^2 |\bar\Psi|^2}{M_P^2}
		+k_4 \frac{|\Phi|^6}{6M_P^4}+\cdots , \\
	W = \kappa \Phi(\Psi \bar \Psi - M^2)+W_0,   \label{Kahler}
\end{gather}
where $W_0 = m_{3/2}M_P^2$ with the gravitino mass $m_{3/2}$,
and $\cdots$ denotes higher dimensional terms.\footnote{
	The constant term may arise dynamically.
	We assume that the field responsible for the $W_0$ already condenses above the energy scale 
	of inflation and it can be regarded as a constant term at scales we are interested in.
}
We keep the non-renormalizable terms in the K\"ahler potential at this stage,
and later we will study the both cases of minimal and non-minimal K\"ahler potential.
We take $\kappa$ real and positive without loosing the generality.
This form of the potential is ensured by assigning $R$-charge $+2$ for $\Phi$ and 0 for $\Psi$ and $\bar \Psi$. 
The scalar potential is calculated from
\begin{equation}
	V_{\rm SUGRA} = e^{K/M_P^2} \left[ K^{i\bar j}D_iW D_{\bar j}\bar W -3\frac{|W|^2}{M_P^2}  \right],
\end{equation}
where $D_iW = W_i + K_i W/M_P^2$, with subscript $i$ denoting the derivative with respective to
the field $i$ and $K^{i \bar j}=K_{i\bar j}^{-1}$.

Suppose that $|\Phi|$ has an initial value much larger than $M$.
Then the $\Psi$ and $\bar\Psi$ have masses of $\kappa |\Phi|$, which is much larger than the Hubble scale
and hence they are stabilized at $\Psi = \bar \Psi = 0$
until $|\Phi|$ rolls down to $M$ where the waterfall behavior turns on.
Since we are interested in the regime before the waterfall occurs, we set $\Psi=\bar\Psi=0$.
Up to the fourth order in $|\Phi|/M_P$, the potential is given by
\begin{equation}
	V_{\rm SUGRA}=\kappa^2 M^4 \left[ 
		1 - k_1 \frac{|\Phi|^2}{M_P^2} + \gamma \frac{|\Phi|^4}{2M_P^4}
	\right]
	+ 2\kappa M^2 m_{3/2}(\Phi+\Phi^*),
\end{equation}
where $\gamma = 1-7k_1/2-3k_4+2k_1^2$.

The inflaton $\Phi$ also receives the potential at one-loop order due to the
mismatch between masses of the scalar and fermion components of $\Psi (\bar \Psi)$.
The Coleman-Weinberg effective potential~\cite{Coleman:1973jx} is given by\footnote{
	There also appears a one-loop correction term of $\sim m_{3/2}^2\Lambda^2$
	in supergravity~\cite{Ferrara:1994kg}.
	But this term does not affect the scalar dynamics, since this is just a constant term and
	smaller than the SUSY breaking contribution $\sim m_{3/2}^2M_P^2$,
	as long as $\Lambda < M_P$. Thus we neglect this term.
}
\begin{equation}
	V_{\rm CW}=\frac{\kappa^4M^4}{32\pi^2}\left[ 
		(x^4+1)\ln \frac{x^4-1}{x^4}+2x^2\ln \frac{x^2+1}{x^2-1}
		+2\ln \frac{\kappa^2 M^2x^2}{\Lambda^2}-3
	\right],
\end{equation}
where $x \equiv |\Phi|/M$. This is approximated as
\begin{equation}
	V_{\rm CW} \simeq \frac{\kappa^4 M^4}{16\pi^2}\ln \frac{\kappa^2 |\Phi|^2}{\Lambda^2},
\end{equation}
in the limit $x\gg 1$.

Thus the total scalar potential is given by the sum of these term: $V = V_{\rm SUGRA} + V_{\rm CW}$.
Writing $\Phi$ as $\Phi = \phi e^{i\theta}/\sqrt{2}$ $(\phi > 0)$, we obtain
\begin{equation}
	V=\kappa^2M^4\left[ 
		1-k_1\frac{\phi^2}{2M_P^2}+\gamma\frac{\phi^4}{8M_P^4}
		+\frac{\kappa^2}{16\pi^2}\ln\frac{\kappa^2\phi^2}{2\Lambda^2}
	\right]
	+2\sqrt{2} \kappa M^2 m_{3/2}\phi \cos\theta.   \label{potential}
\end{equation}
Inflation occurs for some large initial value of $\phi$.
The end of inflation is divided into two cases depending on the parameters.
One is the sudden phase transition at the
waterfall point $\phi_{\rm wf}=\sqrt{2}M$, where inflation terminates due to the
tachyonic instability caused by $\Psi$ and $\bar\Psi$ fields.
The other case is due to the violation of the slow-roll condition, $\eta \sim 1$,
which corresponds to the field value $\phi_{\rm v} \simeq \kappa M_P/(2\sqrt{2}\pi)$.
Thus inflation ends at $\phi = \rm{max}[\phi_{\rm wf},\phi_{\rm v}]$.

Before going into details, we briefly give outlines of the effects of the linear term on the inflaton dynamics.
In the case of the minimal K\"ahler potential ($k_1=0$),
\begin{itemize}
\item The linear term may create a local minimum along the $\phi$ direction for $\pi/2 < \theta < 3\pi/2$,
         and as a result, an angular motion of the inflaton is induced in the complex plane.
	Thus, the inflaton can be trapped by the local minimum even if the inflaton starts with 
	$-\pi/2 < \theta < \pi/2$. For the graceful exit of inflation,  the linear term should be suppressed, 
	 or alternatively, $\theta$ needs to be tuned to be close to $0$.
	For a heavier gravitino, the required tuning on the initial angle $\theta$ becomes severer,
	giving a preference to a lighter gravitino.
\end{itemize}
In the case of the non-minimal K\"ahler potential ($k_1>0$),
\begin{itemize}
\item There is a local minimum due to the $k_1$-term in Eq.~(\ref{potential}) for any $\theta$.
	Thus we must choose the initial position of  $\phi$ 
	so that it does not fall into the local minimum.
\item	The linear term can erase this minimum for a certain range of $\theta$ between $-\pi/2$ and $\pi/2$,
	if the gravitino is heavy enough.
	Thus the required tuning on the initial position $\phi$ becomes weaker for large $m_{3/2}$.
\item However, the heavy gravitino tends to induce larger angular motion of the inflaton,
	and it may drive the inflaton toward the local minimum lying at $\pi/2 < \theta < 3\pi/2$.
	Thus the tuning on the initial angle $\theta$ becomes severer for large $m_{3/2}$.
\item Which of these two effects are important depends on parameters of the model,
	but the tendency is that the lighter gravitino is favored, as will be shown below.
\end{itemize}

\subsection{Fine-tuning issues}   \label{sec:tune}

As outlined above, the initial positions of both the radial $(\phi)$ and angular $(\theta)$ components of the 
inflaton are constrained for successful inflation.
First we discuss the tuning of the radial component, then turn to the angular component.

\subsubsection{Radial component}

Inflation must last at least for about 50 e-folds in order to solve the flatness and horizon problems.
Once the parameters are given, we can calculate the inflaton position at $N\sim 50$.
This should not exceed the Planck scale since for $\phi \gg M_P$ the inflaton
feels an exponential potential which is too steep for inflation to occur, and
hence we limit ourselves to the regime $\phi \ll M_P$ for at least the last $50$ e-foldings.

As a first condition for the initial position of the inflaton, we demand $\phi(N_e) < \phi_i$,
where $\phi(N_e)$ is the inflaton position at the e-foldings $N_e \simeq 50$
and $\phi_i$ is the initial position,
in order for the inflation to last long enough.

In the case of the non-minimal K\"ahler potential, the situation is more complicated.
The inflaton potential (\ref{potential}) in general has a local minimum 
at $\phi_{\rm min}\simeq \sqrt{2k_1}M_P$ for $k_1 > 0$.
This causes a severe initial value problem,
since the inflaton field value must be somehow chosen to evade the local minimum.
This problem is absent for negative $k_1$, but it tends to make the spectral index
blue tilted, which is disfavored by the WMAP observation.
We need to choose the initial position of the inflaton, $\phi_i$,
so that it does not get trapped by the local minimum.
There is a local maximum of the potential at $\phi=\phi_{\rm max} < \phi_{\rm min}$, 
and hence the necessary condition is $\phi_i < \phi_{\rm max}$.
Here $\phi_{\rm max}$ is given by
\begin{equation}
	\phi_{\rm max} \simeq {\rm max}\left\{ \frac{2\sqrt{2}m_{3/2}M_P^2 \cos \theta}{k_1 \kappa M^2},
	~\frac{\kappa M_P}{2\pi \sqrt{2k_1}} \right\}.
\end{equation}
To summarize, we demand $\phi(N_e) < \phi_i < \phi_{\rm max}$, for the non-minimal K\"ahler potential.
The inclusion of the linear term makes both $\phi(N_e)$ and $\phi_{\rm max}$ larger,
and hence it is non-trivial whether the linear term makes the tuning of the initial position severer or not.
We shall see below how severe the tuning is for some parameter choices.

\if 
The condition that the local minimum disappears is expressed by 
\begin{equation}
	V'(\phi_{1}) > 0,
\end{equation}
where $\phi_1$ is a solution of $V''(\phi_1)=0$ and given by
\begin{equation}
	\phi_1^2 = \frac{M_P^2}{2}\left[ 
		\frac{2k_1}{3\gamma}+\sqrt{ \left(\frac{2k_1}{3\gamma} \right)^2 
		+\frac{\kappa^2}{3\gamma \pi^2}  }
	\right].
\end{equation}
In the small $\kappa$ limit, this condition can be rewritten as
\begin{equation}
\begin{split}
	m_{3/2}\cos \theta &> \frac{1}{3\sqrt{3}}\frac{k_1^{3/2}}{\sqrt{\gamma}}\frac{\kappa M^2}{M_P} \\
	& \sim 10^3~{\rm TeV} \left ( \frac{\kappa M^2}{10^{28}~{\rm GeV}^2} \right )
	\left ( \frac{k_1}{0.01} \right )^{3/2} \left(\frac{1}{\gamma}\right)^{1/2}.
	\label{nominimum}
\end{split}
\end{equation}
Thus rather heavy gravitino is necessary.
In the opposite limit $\kappa \gg k_1/\sqrt{\gamma}$, the condition $V'(\phi_{1}) > 0$ is 
always satisfied, hence no tuning is required.
\fi

\subsubsection{Angular component}

The linear term in the inflaton potential (\ref{potential}) induces an angular motion,
which may spoil the inflaton dynamics. In particular, the inflaton trajectory may not reach the 
waterfall point $|\Phi|=M$ due to this angular motion.
In the small $\kappa$ limit, the linear term becomes relatively important for the scalar field dynamics.
In this limit, the equation of motion in the slow-roll regime is written as
\begin{gather}
	3H\dot \phi = 2\sqrt{2}\kappa M^2 m_{3/2}\cos \theta, \\
	3H \dot\theta = 2\sqrt{2}\kappa M^2 m_{3/2}\frac{\sin \theta}{\phi}.   \label{thetadot}
\end{gather}
Assuming $\theta \sim {\rm const.} (0<\theta \lesssim 1)$ and $H\sim \kappa M^2/(\sqrt{3}M_P)$, 
we can solve these equations analytically  and obtain
\begin{equation}
	\phi (N) = \sqrt{2} M\left( 1+2N\frac{m_{3/2}M_P^2}{\kappa M^3}\cos\theta \right),
\end{equation}
where $N$ denotes the e-folding number.
We can estimate the variation of $\theta$ during the last $N$ e-foldings as
\begin{equation}
	\Delta \theta \simeq \tan \theta \ln\left( 1+\delta N \right),   \label{Deltatheta}
\end{equation}
where 
\begin{equation}
	\delta = \frac{2\sqrt{2} m_{3/2}M_P^2 \cos\theta}{\kappa M^3}.
\end{equation}
It is evident that $\delta \sim |W_0\cos \theta / W |$ represents the relative importance of the
constant term in the superpotential. 
In order to avoid the trapping by the local minimum, $\Delta\theta \ll 1$ must be satisfied.
This imposes stringent constraints on the parameter space of hybrid inflation model.
It is rather clear from Eqs.~(\ref{thetadot}) and (\ref{Deltatheta}) that the linear term plays an important role 
no matter what value of $\theta$ is chosen as an initial condition.
In the limit of $\theta \to 0$, the linear term does not induce angular motion, but
it may dominate the dynamics of the radial component.
Thus the slow-roll parameter $\epsilon$ cannot be arbitrary small even in the small $\kappa$ limit.
The inflation scale is then bounded below for generating the density perturbation of the right magnitude.
In the opposite limit $\theta \to \pi/2$, the linear term is relevant for the angular dynamics
and this may spoil the inflaton dynamics in the small $\kappa$ limit.

\begin{figure}
 \begin{center}
   \includegraphics[width=0.5\linewidth]{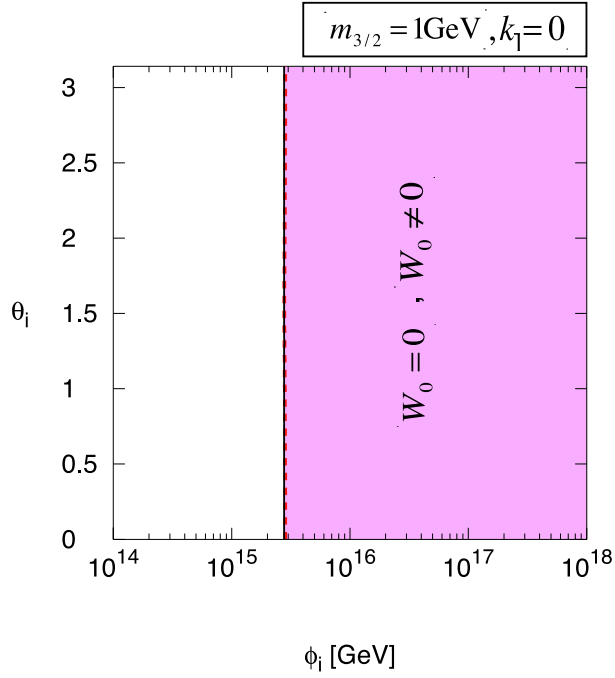}
   \vskip 1cm
   \includegraphics[width=0.5\linewidth]{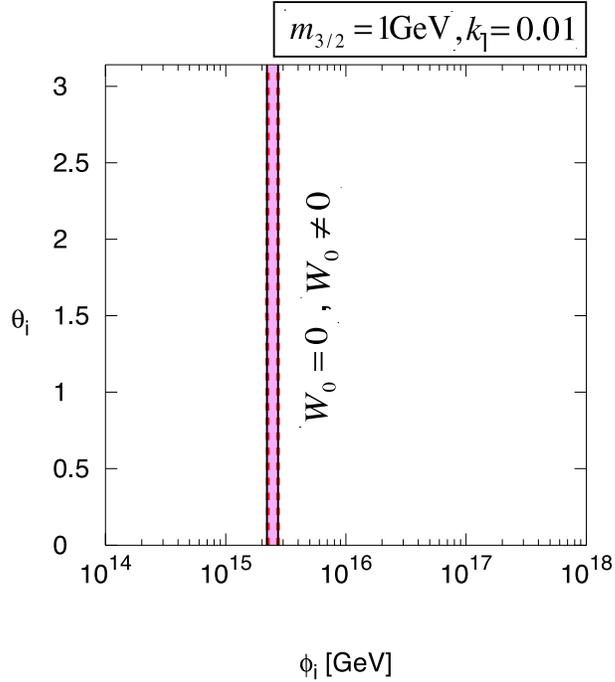}
   \caption{
   	The shaded region indicates the initial position of the inflaton $(\phi_i,\theta_i)$
	where the inflation successfully takes place.
	The region surrounded by the black solid line corresponds to the case 
	of no constant term in the superpotential $W_0=0$, and 
	the region surrounded by the red dashed line corresponds to 
	the more realistic case, $W_0 = m_{3/2}M_P^2$, but they almost coincide with each other.
	We took $\kappa = 10^{-3}, M=10^{14}$~GeV and $m_{3/2}=1$~GeV.
	The upper panel corresponds to the minimal K\"ahler potential $k_1=0$
	and the lower panel corresponds to the nonminimal case with $k_1=0.01$.
    }
   \label{fig:tune_1GeV}
 \end{center}
\end{figure}

\begin{figure}
 \begin{center}
   \includegraphics[width=0.5\linewidth]{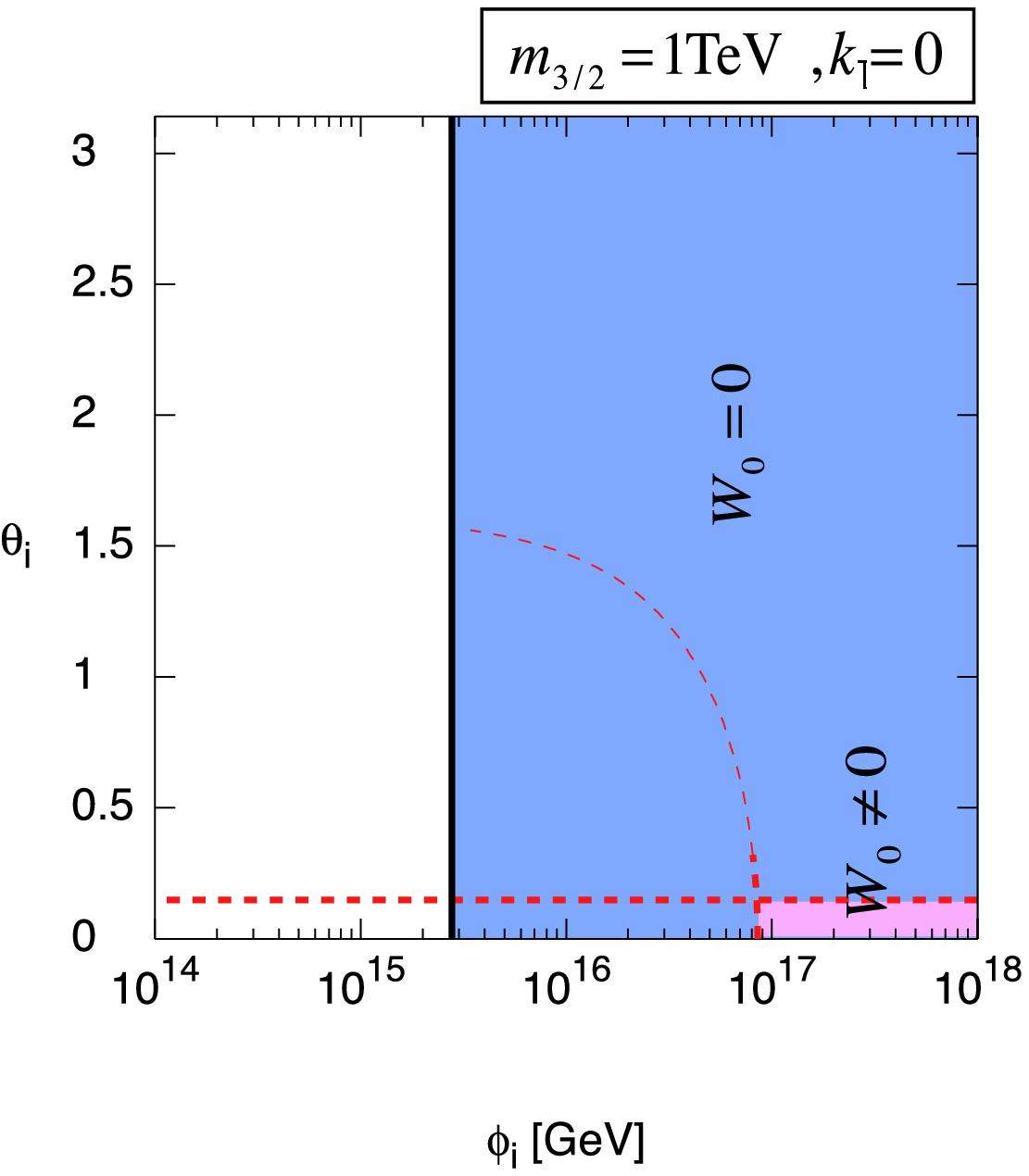}
   \vskip 1cm
   \includegraphics[width=0.5\linewidth]{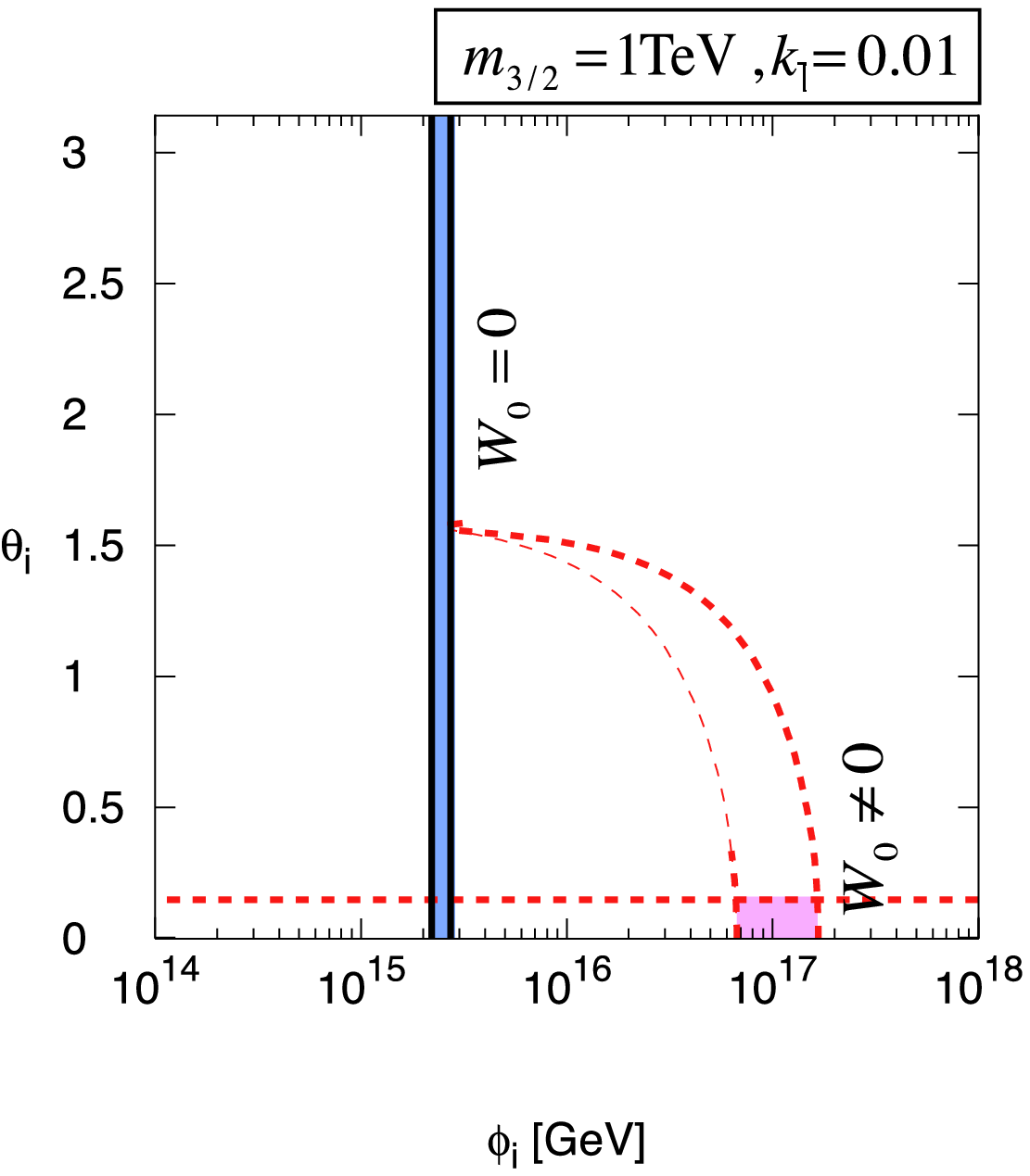}
   \caption{
   	Same as Fig.~\ref{fig:tune_1GeV}, except for $m_{3/2}=1$~TeV.
	The blue shaded region surrounded by the black solid line corresponds to the case 
	of no constant term in the superpotential $W_0=0$, and 
	the pink region surrounded by the red dashed line corresponds to 
	the more realistic case, $W_0 = m_{3/2}M_P^2$.
    }
   \label{fig:tune_1TeV}
 \end{center}
\end{figure}

Figs.~\ref{fig:tune_1GeV} and \ref{fig:tune_1TeV} show the 
initial position of the inflaton $(\phi_i,\theta_i)$ where the inflation successfully takes place. 
Note that here we required that the inflation take place and last for at least $50$ e-folds and
do not impose the WMAP normalization of density perturbation.
We took $\kappa = 10^{-3}, M=10^{14}$~GeV and $m_{3/2}=1$~GeV for Fig.~\ref{fig:tune_1GeV}
and $m_{3/2}=1$~TeV for Fig.~\ref{fig:tune_1TeV}.
The upper panel corresponds to the minimal K\"ahler potential $k_1=0$
and the lower panel corresponds to the nonminimal case with $k_1=0.01$.
The blue shaded region surrounded by the black solid line corresponds to the case 
of no constant term in the superpotential $W_0=0$, and 
the pink region surrounded by the red dashed line corresponds to 
the more realistic case, $W_0 = m_{3/2}M_P^2$.
In Fig.~\ref{fig:tune_1GeV} these two regions almost coincide with each other,
because the effect of the constant term is small enough. As is clearly seen, the initial position 
$\phi_i$ is constrained from below and above for the nonminimal case, which basically comes from
the condition, $M<\phi_i<\phi_{\rm max}$. The presence of the local minimum induced by the
$k_1$-term excludes the region of $\phi > \phi_{\rm max}$, and as a result,  the fine-tuning of
the initial condition becomes severe.  On the other hand,
in Fig.~\ref{fig:tune_1TeV}, the constant term in the superpotential has a significant effect
on the inflaton dynamics. As listed in the end of Sec.~\ref{scalar potential},  $\theta_i$ needs
to be close to zero in order to suppress the angular motion (see Eq.~(\ref{Deltatheta})).
Besides, $\phi_i$ must be large enough in order for the inflation to last for more than $50$ e-folds.


\section{Constraints on hybrid inflation models}  \label{sec:const}

\subsection{Successful inflation}    \label{sec:suc}

In the previous section we have discussed the fine-tuning issues of  the initial condition.
Actually, however, there are parameter regions where successful inflation cannot occur
regardless of the choice of the initial condition.

First, we demand $\phi (N_e) < M_P$ as the necessary condition for successful inflation.
In other words, unless this inequality is satisfied, 
inflation cannot amount to 50 e-foldings even if the initial inflaton position is set to be around the Planck scale.
This corresponds to the limit where the lower bound on $\phi_i$ in 
Figs.~\ref{fig:tune_1GeV} and \ref{fig:tune_1TeV} goes to $M_P$ and the allowed region vanishes. 


In the case of the non-minimal K\"ahler potential with positive $k_1$,
there appears  local minimum and maximum at $\phi = \phi_{\rm min}$ and $\phi_{\rm max}$,
respectively, as explained in the previous section.
In order for inflation to take place successfully, we need $M < \phi_{\rm max}$,
since otherwise the waterfall field always triggers the end of inflation.
In the limiting case $ \phi_{\rm max} = M$, the lower and upper bound on $\phi_i$
in Figs.~\ref{fig:tune_1GeV} and \ref{fig:tune_1TeV} cross with each other and allowed region disappears.

\subsection{Density perturbation}

The inflaton quantum fluctuations seed the large-scale structure of the Universe,
and the WMAP observation of the CMB anisotropy indicates 
$\Delta_{\zeta}^2 \simeq 2.4\times 10^{-9}$ in terms of the 
dimensionless power spectrum of the curvature perturbation at the pivot scale 
$k = 0.002~{\rm Mpc}^{-1}$~\cite{Komatsu:2010fb}.
It is related to the inflaton potential through the relation
\begin{equation}
	\Delta_{\zeta}^2 = \frac{V}{24\pi^2 M_P^4 \epsilon},\label{norm}
\end{equation}
where $\epsilon = (M_P^2/2)(V'/V)^2$ is the slow-roll parameter
and the potential and slow-roll parameter should be evaluated when the scale of interest 
exits the horizon, $N_e =  50 \sim 60$.

The scalar spectral index, $n_s$, is calculated from $n_s = 1-6\epsilon +2\eta$,
where $\eta = M_P^2V''/V$ is another slow-roll parameter and estimated as
\begin{equation}
	\eta \simeq -k_1 + \frac{3\gamma}{2}\frac{\phi^2}{M_P^2} 
	- \frac{\kappa^2}{8\pi^2}\frac{M_P^2}{\phi^2}.
\end{equation}
Here we have approximated as $V \simeq \kappa^2 M^4$.
As is obvious, the $k_1$ term gives a constant contribution to the $\eta$ parameter.
For inflation to occur, $|\eta| \ll 1$ is required.
If $k_1$ is positive, it tends to make the spectral index red tilted $(n_s < 1)$~\cite{BasteroGil:2006cm}.
Therefore, the introduction of positive small $k_1 (\sim 0.01)$ is well-motivated,
but it simultaneously creates a local minimum of the potential which makes the inflaton
dynamics nontrivial, as explained before.
We consider two cases, $k_1=0$ and $k_1=0.01$ as reference values in the following analysis.

\subsection{Cosmic strings}

After inflation, $\Psi$ and $\bar \Psi$ obtain a large VEV $M$, which breaks the $U(1)_{B-L}$ symmetry,
and hence cosmic strings are formed.
The tension of the cosmic string, $\mu$, is given by 
\begin{equation}
	\mu = 2\pi M^2 \gamma(\kappa),
\end{equation}
where $\gamma$ is a weakly varying function of $\kappa$,
which takes the value of $\mathcal O(0.1)$~\cite{Jeannerot:2005mc}.
Cosmic strings constantly generate density perturbations in a way different from the
inflationary adiabatic perturbation, and such a contribution is restricted from observations
\cite{Hill:1987ye,Battye:2006pk}.
The recent bound reads $G\mu \lesssim (2-7)\times 10^{-7}$
from the CMB and matter power spectrum data~\cite{Battye:2010xz}.
In this paper we demand $G\mu < 5\times 10^{-7}$ as a representative value.

\subsection{Gravitino problem}

There are two sources of the gravitino production.\footnote{
	We do not consider the cosmological effects of moduli, which may appear in 
	the compactification of the extra dimensions in string theory.
	If they exist, cosmological constraints would become much more stringent~\cite{Banks:1993en}. 
}
One is the scattering processes between particles in thermal bath at the reheating.
The gravitino abundance is calculated as~\cite{Bolz:2000fu,Kawasaki:2004qu}
\begin{equation}
\begin{split}
	Y_{3/2}^{\rm (TP)} \simeq 2\times 10^{-12}\left ( 1+\frac{m_{\tilde g}^2}{3m_{3/2}^2} \right )
	\left ( \frac{T_{\rm R}}{10^{10}~{\rm GeV}} \right ),
	\label{thY}
\end{split}
\end{equation}
where $T_{\rm R}$ is the reheating temperature after inflation, and $m_{\tilde g}$ denotes the gluino mass.

The other source of the gravitino is the decay of the inflaton~\cite{Kawasaki:2006gs,Endo:2007ih,Asaka:2006bv,Endo:2006tf}. 
The non-thermal gravitino production inevitably occurs if the inflaton has a non-vanishing VEV, or more precisely, if $\la K_\phi \ra$ does not
vanish. Importantly, the gravitino abundance is inversely proportional to the reheating temperature, in contrast
to the thermal scattering (\ref{thY}). Therefore, the non-thermal gravitino production gives a tight 
constraint on the inflation models, when combined with the thermal scattering production.
The abundance is estimated as~\cite{Kawasaki:2006gs,Endo:2007ih}
\begin{equation}
	Y_{3/2}^{\rm (NTP)}\simeq 
	7\times 10^{-11}\left ( \frac{g_*}{200} \right )^{-1/2}
	\left ( \frac{T_{\rm R}}{10^6~{\rm GeV}} \right )^{-1}
	\left ( \frac{\langle \phi \rangle}{10^{15}~{\rm GeV}} \right )^2
	\left ( \frac{m_\phi}{10^{12}~{\rm GeV}} \right )^2,
\end{equation}
for $m_\phi \lesssim \Lambda$, and
\begin{equation}
	Y_{3/2}^{\rm (NTP)}\simeq 
	9\times 10^{-13}\xi \left ( \frac{g_*}{200} \right )^{-1/2}
	\left ( \frac{T_{\rm R}}{10^6~{\rm GeV}} \right )^{-1}
	\left ( \frac{\langle \phi \rangle}{10^{15}~{\rm GeV}} \right )^2
	\left ( \frac{m_\phi}{10^{12}~{\rm GeV}} \right )^2,
\end{equation}
for $m_\phi \gtrsim \Lambda$, 
where $\langle \phi \rangle$ and $m_\phi$ denote the VEV and mass of the inflaton, respectively,
$\Lambda$ is the mass of the SUSY breaking field,
and $\xi$ is a model-dependent constant of order unity.
In this paper we take $\Lambda = \sqrt{m_{3/2}M_P}$ for the sake of concreteness, but this choice
does not much affect the results. As can be seen from the above expressions, more gravitinos are produced
for a larger inflaton VEV and a heavier inflaton mass. Therefore a high-scale inflation model tends to be
disfavored by the inflaton-induced gravitino problem.

Thus, the total gravitino abundance  is given by $Y_{3/2}=Y_{3/2}^{\rm (TP)}+Y_{3/2}^{\rm (NTP)}$.
In the hybrid inflation model, 
$\langle \phi \rangle = M$ and $m_\phi = \sqrt{2}\kappa M$, and the gravitino problem
gives stringent constraints on the parameter space $\kappa$ and $M$.
In the following, in order to derive conservative bounds on the parameter space,
we set $T_{\rm R}$ to be the maximally allowed value from thermal production contribution
so that the non-thermal contribution from the inflaton decay is minimized~\cite{Kawasaki:2006gs}.

The constraint depends on the gravitino mass.
If the gravitino is the LSP, as is usually realized in 
gauge-mediated SUSY breaking models~\cite{Dine:1993yw},
its abundance is bounded from above so that it does not exceed the
present dark matter abundance. This reads
\begin{equation}
	m_{3/2}Y_{3/2} \lesssim 4\times 10^{-10}~{\rm GeV},
\end{equation}
for $m_{3/2} \lesssim 100~{\rm GeV}$.
In the gravity-mediated SUSY breaking scenario, the gravitino mass is expected to be around 1~TeV
and it decays after BBN begins.
In order not to spoil the success of BBN, the gravitino abundance is severely constrained.
The constraint sensitively depends on the gravitino mass, and typical upper bound is~\cite{Kawasaki:2004qu}
\begin{equation}
	m_{3/2}Y_{3/2} \lesssim 10^{-14}~{\rm GeV},
\end{equation}
for $m_{3/2}\sim 1$~TeV.
In anomaly-mediated SUSY breaking scenario~\cite{Randall:1998uk}, 
the gravitino mass is much heavier, and hence it decays well before BBN begins.
However, the gravitino decay non-thermally produces LSPs 
(often the Wino-like neutralino in the anomaly mediation),
whose abundance must not exceed the dark matter abundance.
Thus the constraint reads
\begin{equation}
	m_{3/2}Y_{3/2} \lesssim 4\times 10^{-10}~{\rm GeV}\left ( \frac{m_{3/2}}{m_{\rm LSP}} \right ),
\end{equation}
for $m_{3/2} \gg 10$~TeV.\footnote{
	In this expression we ignored self-annihilation of the LSP after the non-thermal production.
	If the annihilation cross section is fairly large 
	or if the gravitino decay temperature is sufficiently high,
	it can reduce the resulting LSP abundance and the bound can be loosen. 
}
In the following analysis we make use of the anomaly-mediation relation
between masses of the Wino-like neutralino LSP and the gravitino, 
$m_{\rm LSP} = (g_2^2/16\pi^2)m_{3/2}$ where $g_2$ denotes the SU(2) coupling constant.

\subsection{Constraints on parameter space} 

\subsubsection{Minimal K\"ahler potential} 

First let us consider the minimal K\"ahler potential, i.e., $k_1=k_2=k_3=k_4=0$ in Eq.~(\ref{Kahler}).
The Hubble-induced mass term for the inflaton is cancelled in the supergravity potential.

\begin{figure}
 \begin{center}
   \includegraphics[width=0.6\linewidth]{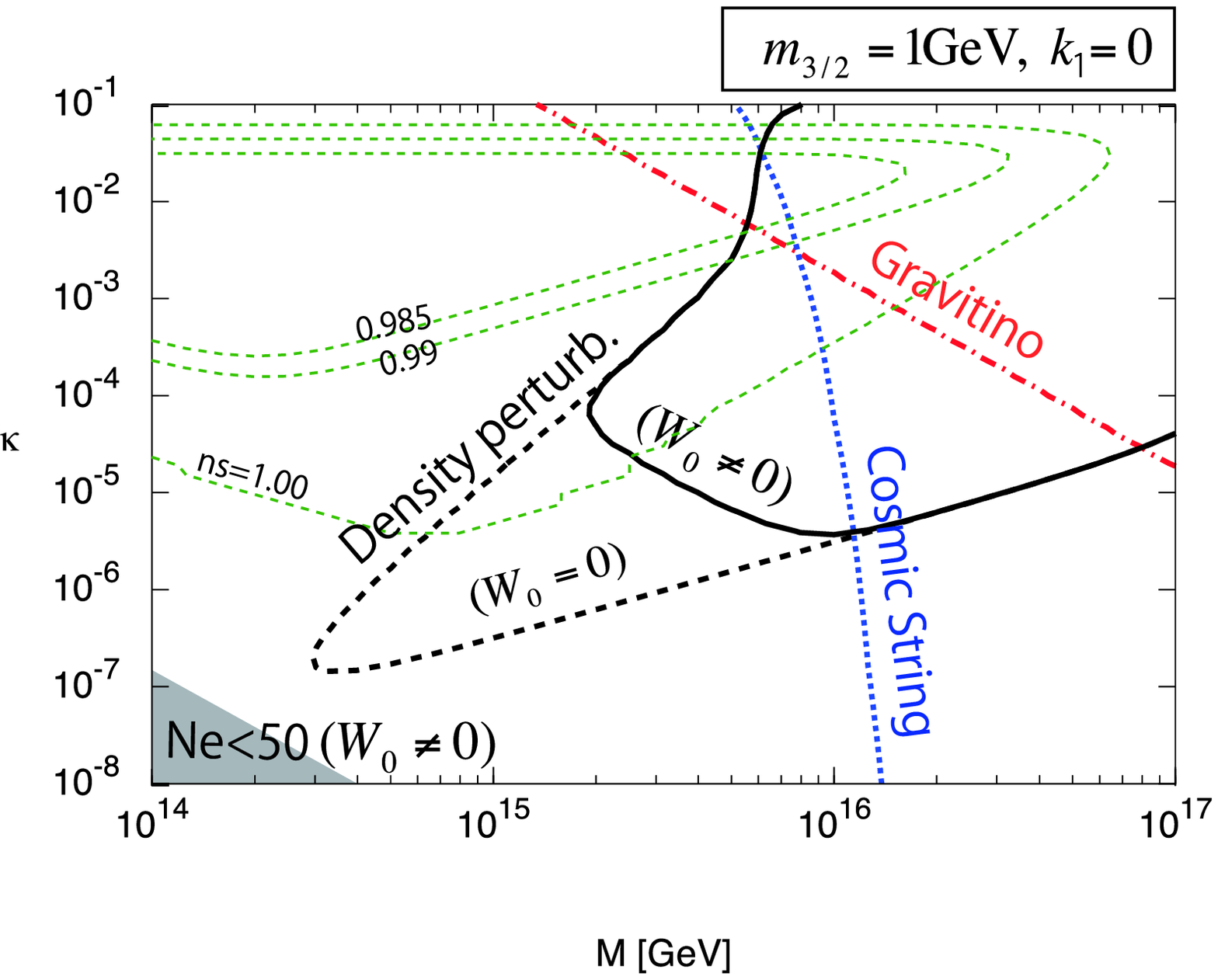}
   \vskip 1cm
   \includegraphics[width=0.6\linewidth]{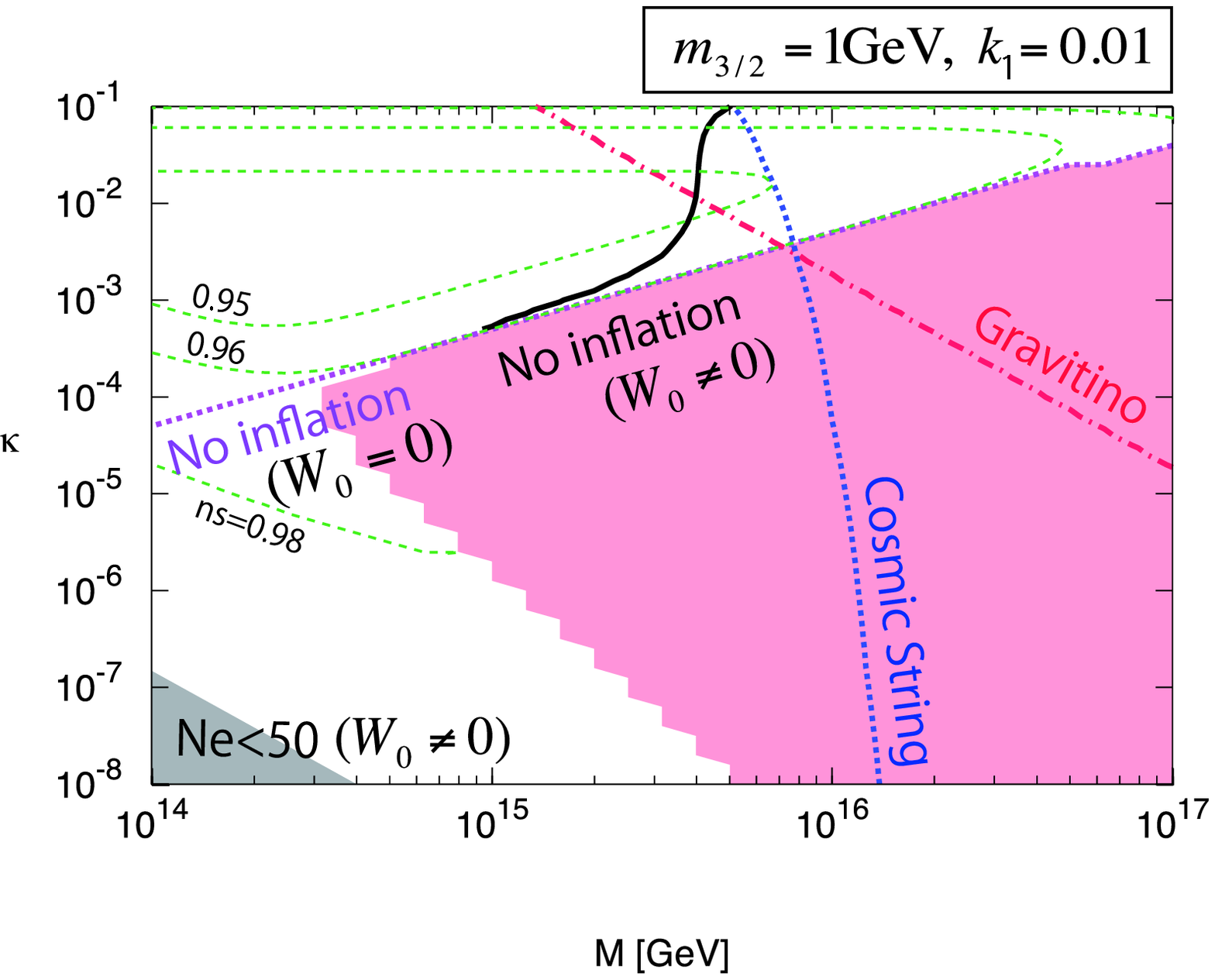}
   \caption{
   	Constraints on $(M,\kappa)$ plane for $m_{3/2}=1$~GeV, $\theta = 0$ 
	and $k_1 = 0$ (top) and $k_1=0.01$ (bottom).
	Along the black solid (dashed) line, the density perturbation of the right magnitude is generated
          for $W_0=m_{3/2}M_P^2$ $(W_0=0)$.
	The blue dotted line shows the upper bound on $M$ from the constraint on cosmic strings.
	Above the red dot-dashed line, gravitinos are overproduced.
	In the shaded region at the bottom-left corner of the panels, the linear term disturbs the flatness of the
	potential and inflation does not last long enough, i.e., $\phi(N_e) > M_P$,
	for the case of $W_0=m_{3/2}M_P^2$.
	In the bottom panel, the purple shaded region is excluded for $W_0=m_{3/2}M_P^2$,
	since inflation does not occur, i.e., $M > \phi_{\rm max}$.
	For comparison, we have also shown the region where inflation does not occur for the case of $W=0$
	by the purple dotted line.
	Contours of $n_s$ are also plotted by thin green dashed lines.
    }
   \label{fig:1GeV}
 \end{center}
\end{figure}

\begin{figure}
 \begin{center}
   \includegraphics[width=0.6\linewidth]{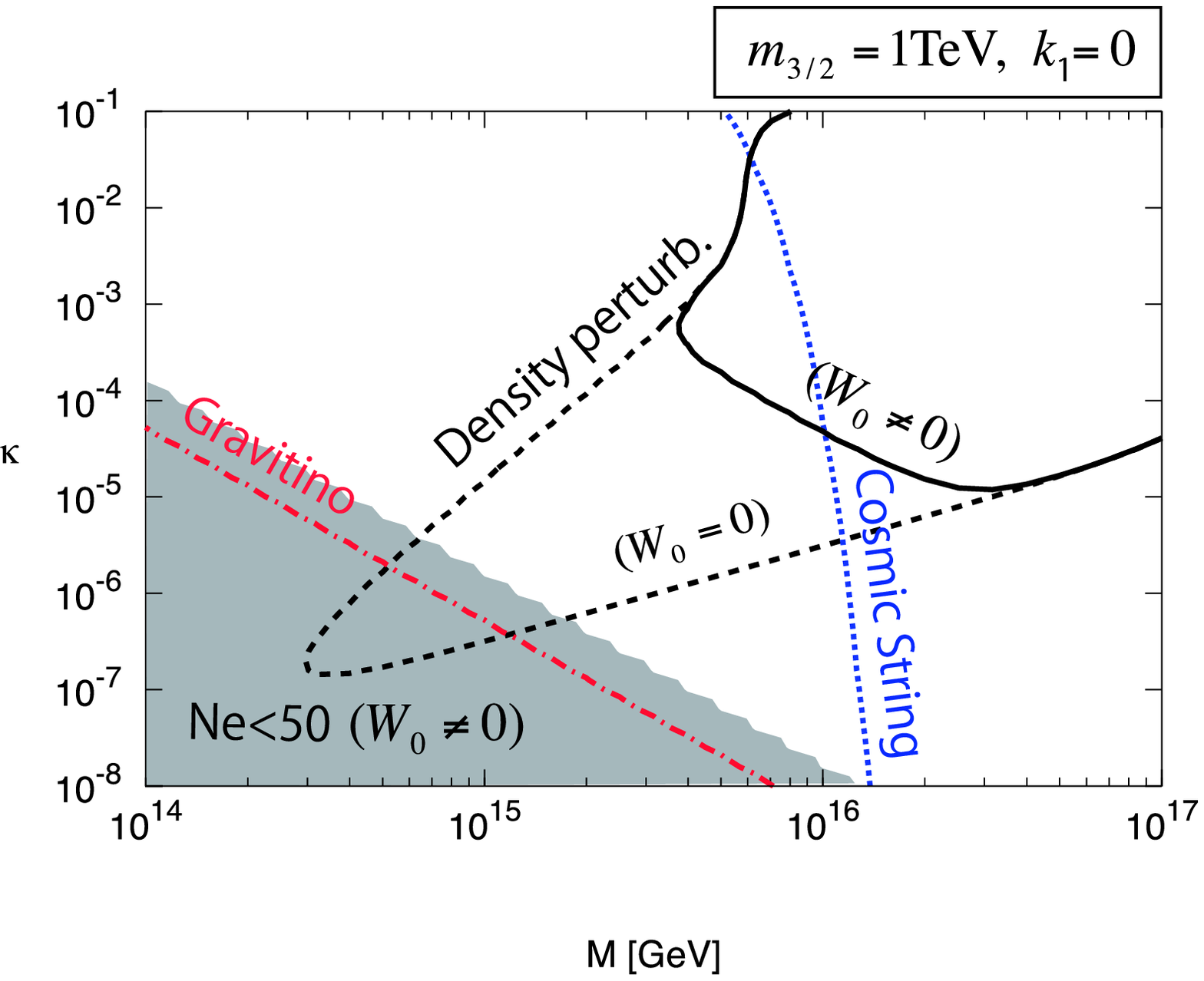}
   \vskip 1cm
   \includegraphics[width=0.6\linewidth]{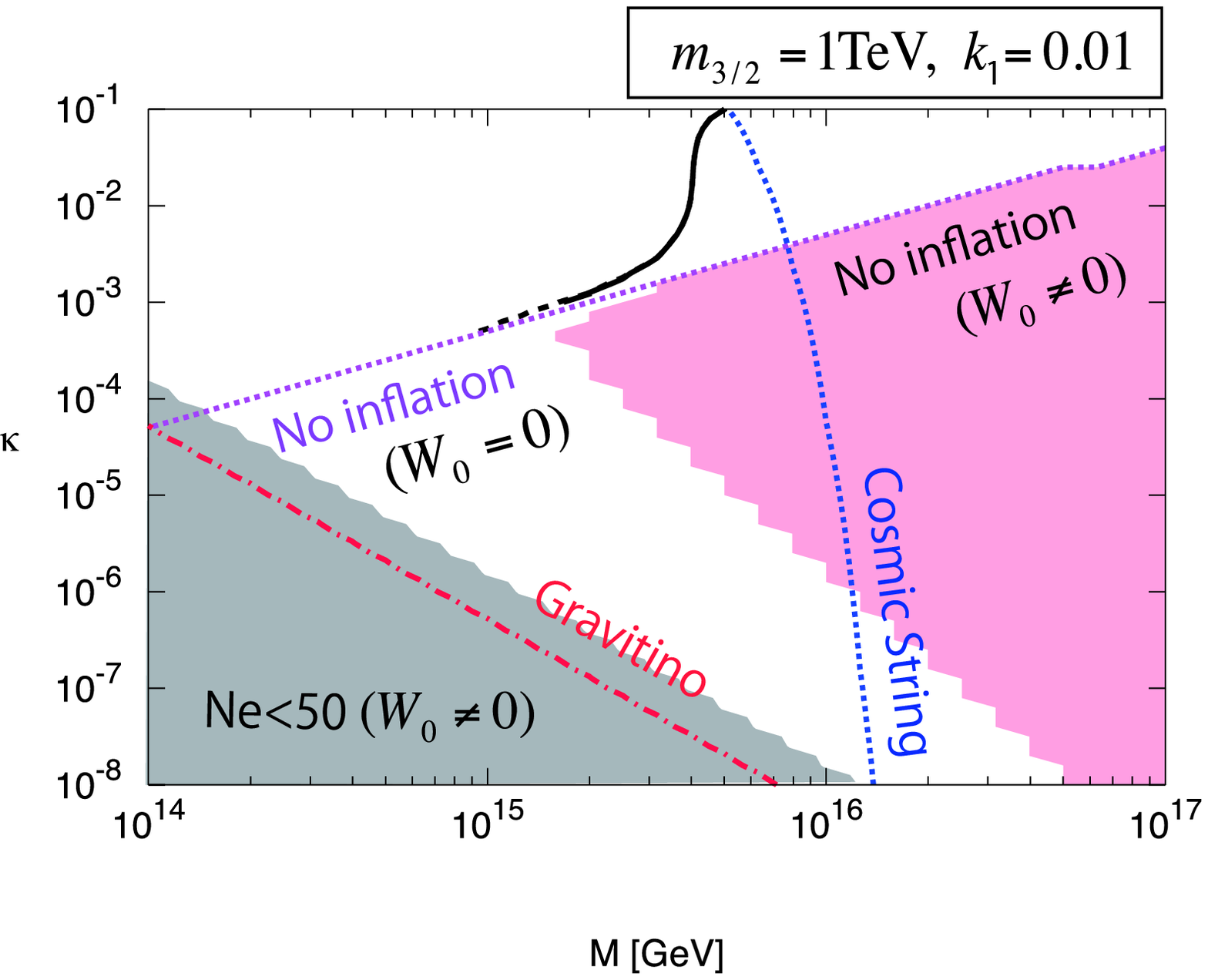}
   \caption{ 
   	Same as Fig.~\ref{fig:1GeV}, but for $m_{3/2}=1$~TeV.
    }
   \label{fig:1TeV}
 \end{center}
\end{figure}

\begin{figure}
 \begin{center}
   \includegraphics[width=0.6\linewidth]{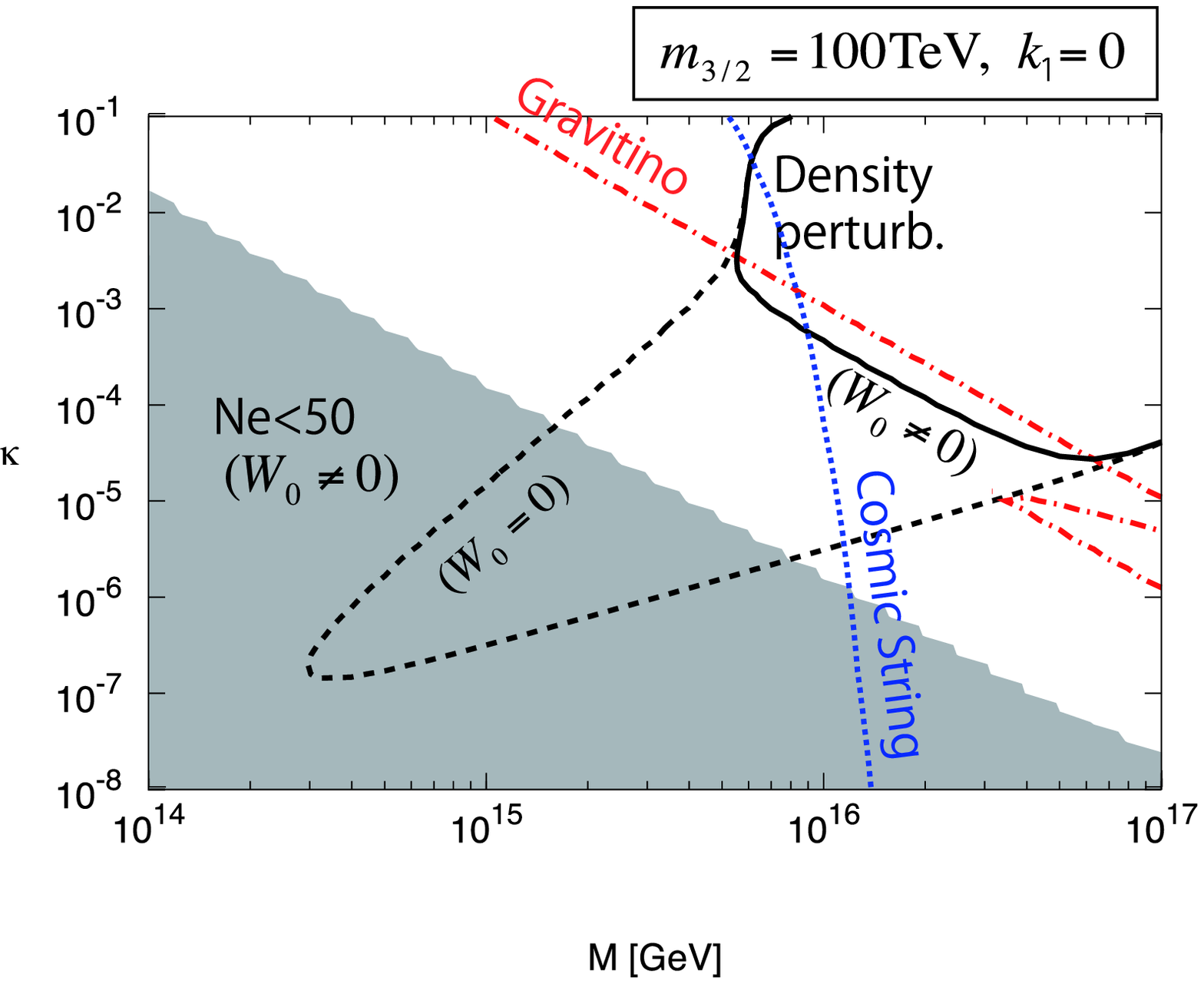}
   \vskip 1cm
   \includegraphics[width=0.6\linewidth]{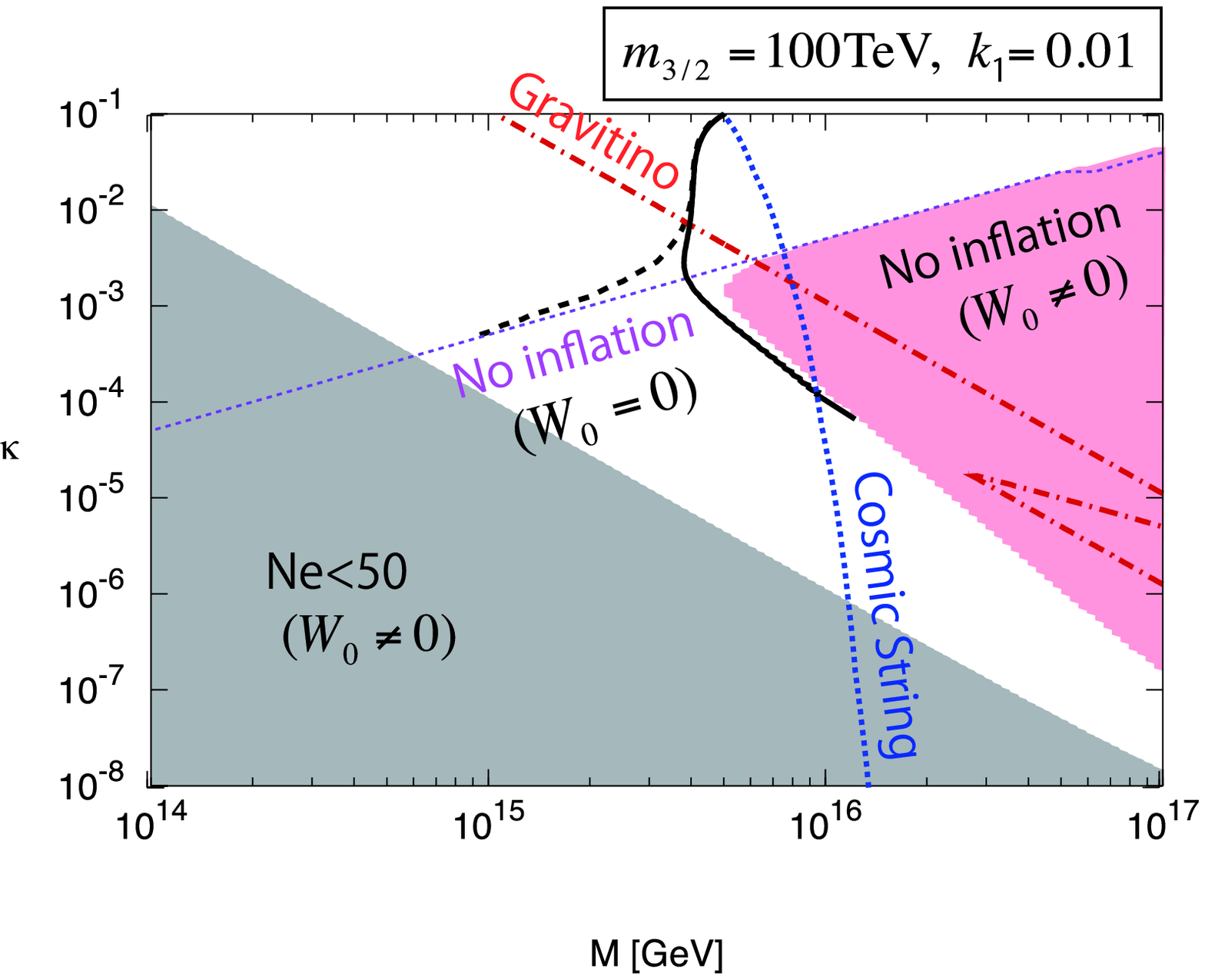}
   \caption{ 
   	Same as Fig.~\ref{fig:1GeV}, but for $m_{3/2}=100$~TeV.
    }
   \label{fig:1000TeV}
 \end{center}
\end{figure}

The constraints on $(M,\kappa)$ plane are shown in Figs.~\ref{fig:1GeV}-\ref{fig:1000TeV}
for $m_{3/2}=1$~GeV, 1~TeV and $100$~TeV,  respectively.
The upper panel in each figure corresponds to the case of the minimal K\"ahler potential.
Here the initial angle of the inflaton is chosen to be $\theta = 0$
so that the angular motion is not induced by the linear term. 
For $\theta \ne 0$, the viable parameter regions shrink.
For comparison, we have shown the case of no constant term in the superpotential, $W_0=0$.
Along the black solid (dashed) line, the density perturbation of the correct magnitude is generated
for $W_0=m_{3/2}M_P^2$ $(W_0=0)$.
The blue dotted line shows the upper bound on $M$ from the constraint on the cosmic string
Above the red dot-dashed line, gravitinos are overproduced.
In the shaded region at the bottom-left corner of the figure, the linear term disturbs the flatness of the
potential and inflation does not last long enough, i.e., $\phi(N_e) > M_P$.

The contours of the scalar spectral index $n_s$ are also plotted by thin green dashed lines.
It should be noted that $n_s$ is calculated without including the contribution from cosmic strings
to the density perturbation.
Thus in the region where the cosmic string effect is significant, 
the actual spectral tilt of the observed density perturbation may not be directly related to $n_s$ plotted here.
In the minimal K\"ahler potential, the scalar spectral index is close to unity
for most of the parameter space. This is because the inflaton potential is relatively flat and the inflaton 
does not move much and stays around the waterfall point during the last 50 e-foldings.
An exception is for a relatively large $\kappa$ region $(\kappa \gtrsim 0.01)$
where the one-loop effect is important.
In that region, the spectral index can be around $0.98$ and 
the consistency with current observations is better.
However, the region is on the boundary of the cosmic string bound, and significant 
cosmic string contribution to the density perturbation may ameliorate the situation.
Therefore, search for the cosmic strings in the CMB anisotropies and the gravity waves 
may confirm or disfavor the model.


\subsubsection{Non-minimal K\"ahler potential} 

Next, we investigate the case of the non-minimal K\"ahler potential.
The most important term is the $k_1$-term in Eq.~(\ref{Kahler}), since it 
directly affects the scalar spectral index. In fact, $k_1 \sim 0.01$ gives a better fit
to the WMAP observation of the spectral index $n_s = 0.963\pm 0.012$ (68\% C.L.).
However, the $k_1$ term generically creates a local minimum in the inflaton potential.
Thus we need to fine-tune the initial position of the inflaton to be smaller than the local maximum.
But even this tuning is not always sufficient, as explained in Sec.~\ref{sec:suc}.
If the position of the local maximum $\phi_{\rm max}$ is smaller than $M$,
inflation does not take place for any choice of the initial position.

The constraints on $(M,\kappa)$ plane are shown in the bottom panels of Figs.~\ref{fig:1GeV}-\ref{fig:1000TeV}
in the case of the non-minimal K\"ahler potential with $k_1=0.01$
for $m_{3/2}=1$~GeV, 1~TeV and $100$~TeV, respectively.
The initial angle of the inflaton is chosen to be $\theta = 0$
so that the angular motion is not induced by the linear term.
The meanings of lines are same as the top panels.
The purple shaded region is excluded for $W_0=m_{3/2}M_P^2$,
since inflation does not occur, i.e., $M > \phi_{\rm max}$.
For comparison, we have also shown the region where inflation does not occur for the case of $W=0$
by the purple dotted line; the region below the line is excluded. 
It should be noted that in the non-shaded region, some amount of fine-tuning concerning the 
initial position of the inflaton is necessary in the sense of Sec.~\ref{sec:tune} (see Figs.~\ref{fig:tune_1GeV} and \ref{fig:tune_1TeV}).

There remains a consistent parameter region
for relatively large $\kappa \sim 10^{-3}$ for $m_{3/2} \lesssim 1$~GeV.
Around this region, the scalar spectral index is close to 0.96
because of the $k_1$-term in the scalar potential, and hence the fit to the observation is better.

\subsubsection{Comparison between different gravitino masses} 
Let us focus on the two constraints: the gravitino problem and the effect of the
linear term. As emphasized in Ref.~\cite{Kawasaki:2006gs}, the gravitino problem,
especially when non-thermal production is taken into account, excludes quite a few number
of inflation models in the gravity mediation. So, the gravitino problem suggests that
the gravitino mass is either heavy ($\gtrsim O(10)$\,TeV) or light ($\lesssim O(1)$\,GeV),
for which the cosmological effect of the gravitino is mild. Indeed, in the case of the hybrid inflation 
model, there are allowed regions for heavy or light gravitinos, where the inflaton mass
is lighter than about $\GEV{14}$. In other words, one could avoid the gravitino bound if
$\kappa$ and/or $M$ is small enough. 

However, the effect of the linear term was not taken into account in the above analysis.
The linear term is more important for a heavier gravitino, as one can see from the figures that
the gray shaded (bottom-left corner) regions become larger for a heavier gravitino.
In particular, the black dashed lines, on which the density perturbation of the right magnitude
would be generated for $W_0 = 0$, are significantly modified to the solid lines, especially for $m_{3/2} = 100$TeV.
This is because, due to the additional tilt induced by the linear term, the inflation scale must be
larger to generate the density perturbation with the same magnitude (see Eq.~(\ref{norm})).
Such modification of the parameters produce a tension with the gravitino problem
especially for a heavy gravitino. Thus, we conclude that, combining the gravitino problem and
the effect of the linear term, a light gravitino is preferred in the hybrid inflation model.

\section{Nonthermal leptogenesis}  \label{sec:lepton} 

Finally let us consider the possibility that the inflaton decays into right-handed neutrinos
through the interaction (\ref{PNN}) and subsequent decay of the right-handed neutrinos
generates a lepton asymmetry, which is transformed into a baryon asymmetry
through the sphaleron process~\cite{Fukugita:1986hr}.

The abundance of the right-handed neutrino, $N_1$, produced non-thermally by the inflaton decay,
is given by~\footnote{
	We consider only the lightest right-handed neutrino, $N_1$, for simplicity.
	This can be justified by demanding $N_2$ and $N_3$ heavier than the inflaton 
	and not produced by the inflaton decay.
}
\begin{equation}
	\frac{n_{N_1}}{s} = \frac{3}{2}B_r \frac{T_{\rm R}}{m_\phi},
\end{equation}
where $B_r$ is the branching ratio of the inflaton decay into $N_1$ pair.
We assume the mass of $N_1$ is larger than the reheating temperature: $m_{N_1} > T_{\rm R}$.
In this case, the right-handed neutrinos are not thermalized after inflation
and we do not need to take account of the wash-out effects of the lepton asymmetry.
The CP-asymmetric decay of $N_1$ produces net baryon number after sphaleron process, 
given by~\cite{Asaka:1999yd}
\begin{equation}
\begin{split}
	\frac{n_B}{s} 
	\simeq 1\times 10^{-10} B_r
	\left( \frac{T_{\rm R}}{10^6~{\rm GeV}} \right)
	\left( \frac{m_{N_1}}{m_\phi} \right)
	\left( \frac{m_{\nu 3}}{0.05~{\rm eV}} \right)   
	\delta_{\rm eff},     \label{baryon}
\end{split}
\end{equation}
where $\delta_{\rm eff}$ is an $\mathcal O(1)$ parameter which represents the effective CP phase.
Here we have assumed that the right-handed neutrino decays immediately after the inflaton decay.
This is justified unless the right-handed neutrino is extremely lighter than the inflaton.

In the present model, $m_\phi = \sqrt{2}\kappa M$ and $m_{N_1} = y_{11}M$.
In order for the inflaton decay into $N_1$'s to be kinematically accessible,
$\kappa > 2y_{11}$ is imposed.
We can always take $y_{11}$ to be not so far from $\kappa$, so that the produced baryon number
is not suppressed (Eq.~(\ref{baryon})).
After inflation, the scalar fields $\Phi$ and $\Psi_{\rm m}=(\Psi+\bar\Psi)/\sqrt{2}$ have
same masses $m_\phi = \sqrt{2}\kappa M$ and they are maximally mixed with each other
\cite{Kawasaki:2006gs}.
The decay rate of $\Psi_{\rm m}$ into two $N_1$'s is given by
\begin{equation}
	\Gamma(\Psi_{\rm m} \to N_1N_1) = \frac{1}{64\pi}y_{11}^2 m_\phi 
	\left(1- \frac{4m_{N_1}^2}{m_\phi^2} \right)^{3/2}.
\end{equation}
The decay rate of $\Phi$ into sneutrinos, $\Phi \to \tilde N_1\tilde N_1$, is almost the same
\cite{Endo:2006ix}.
Assuming these are the dominant decay process of the inflaton ($B_r = 1$), 
the reheating temperature is estimated as
\begin{equation}
	T_{\rm R} \simeq 3\times 10^7~{\rm GeV}
	\left( \frac{\kappa}{10^{-5}} \right)^{3/2}
	\left( \frac{M}{10^{15}~{\rm GeV}} \right)^{1/2}
	\left( \frac{m_{N_1}}{m_\phi} \right).       \label{TR}
\end{equation}
This cannot be much larger because of the gravitino problem.
On the other hand,
as is clear from Eq.~(\ref{baryon}), $T_{\rm R}\gtrsim 10^6$~GeV is required for reproducing the
observed amount of baryon asymmetry of the Universe even for $m_{N_1}\sim m_\phi$.

In the case of gauge mediation, thermal gravitino production implies 
$m_{3/2}\gtrsim O(10)$\,MeV from $T_{\rm R} \gtrsim 10^6~$GeV for successful leptogenesis.
One can see from Fig.~\ref{fig:1GeV} that there is a parameter region where
the non-thermal leptogenesis works. For instance, in the case of the minimal K\"ahler potential,
we find that  $\kappa \sim 10^{-5}$, $M\sim 3\times 10^{15}~$GeV and $m_{N_1}\sim 0.1m_\phi$ are
viable parameter choice for $m_{3/2}=1$~GeV (see top panel of Fig.~\ref{fig:1GeV}).

In the case of gravity mediation where $m_{3/2} = 100$\,GeV $\sim 1$\,TeV,
the upper bound on the reheating temperature reads $T_{\rm R}\lesssim 10^{6-7}$~GeV.
Thus we need $m_{N_1}\sim m_\phi$ for correct amount of baryon asymmetry (\ref{baryon}).
Comparing it with Eq.~(\ref{TR}), smaller $\kappa (< 10^{-5})$ or $M (<10^{15}$~GeV) is required,
but this is not compatible with the density perturbation condition 
(see the top panel of Fig.~\ref{fig:1TeV}).
In the non-minimal K\"ahler case, the situation becomes worse (see the bottom panel of Fig.~\ref{fig:1TeV}).

In the case of anomaly-mediation where $m_{3/2}\gtrsim 100$~TeV,
the upper bound on the reheating temperature is loosened as $T_{\rm R}\lesssim 10^{10}$~GeV.
On the other hand, rather large $\kappa$ and $M$ is needed for satisfying the density perturbation condition
(see the top panel of Fig.~\ref{fig:1000TeV}).
For instance, in the case of the minimal K\"ahler potential, we find 
$\kappa \sim 10^{-3}$, $M\sim \GEV{16}$ and $m_{N_1}\lesssim 0.1m_\phi$ can satisfy the bound
on the reheating temperature. 
The situation is similar in the non-minimal K\"ahler case 
(see the bottom panel of Fig.~\ref{fig:1000TeV}).

\section{Conclusions and discussion}   \label{sec:conc}

We have reexamined the hybrid inflation model in supergravity,  and searched for a viable parameter region,
particularly taking account of the non-thermal gravitino production and the effect of the linear term in the superpotential. 
We have found that the effect of the linear term on the inflaton dynamics as well as the fine-tuning of the initial condition
becomes significant for a heavy gravitino mass, giving a clear preference to a light gravitino. Importantly, this constraint
disfavors a heavy gravitino ($\gtrsim 100$\,TeV), which is otherwise allowed by the gravitino problem. 
Remember that the non-thermal gravitino production restricts the inflation model parameters especially for
unstable weak-scale gravitino, allowing for either heavy or light gravitino. 
In either case, it is likely that the predicted tension of the cosmic string is close to the present upper bound,
and may be within the reach of future detection.
We also confirmed that the non-thermal leptogenesis could work for $m_{3/2}=O(10)$\,MeV $\sim 1$\,GeV,
and $m_{3/2} \gtrsim O(10)$\,TeV. For $m_{3/2} \lesssim 16$\,eV~\cite{Viel:2005qj}, there is no upper bound on the
reheating temperature, and so, thermal leptogenesis is possible. 
 

In order to avoid the above tight constraints on the heavy gravitino, we need some extension or non-trivial cosmological assumptions: 
entropy production to dilute the gravitino, 
curvaton for producing the density perturbation, etc. 

Some comments are in order.
In this paper we have concentrated on the hybrid inflation model in supergravity.
One may wonder whether the constant term in the superpotential has drastic effects on
inflationary dynamics or not in other inflation models.
In the new inflation model with superpotential $W = \phi (\mu^2 - \phi^n/M^{n-2})$~\cite{Izawa:1996dv}
we do not need to add a constant term in the superpotential :
it is automatically generated after inflation ends and the inflaton relaxes to the potential minimum.
In the variant type of new inflation model like $W = \phi (\mu^2 - \psi^n/M^{n-2})$~\cite{Asaka:1999yd},
where $\psi$ takes a roll of the inflaton, the constant term should be added
in order to make sure that the cosmological constant is almost zero in the present Universe.
In this case, under a reasonable assumption, the $\phi$ field settles at the origin $\phi = 0$ during inflation,
and hence all the terms in the scalar potential 
which are included by the constant term are highly suppressed.
Hence we expect no drastic effect of the constant term in the superpotential.
In the chaotic inflation model in Ref.~\cite{Kawasaki:2000yn},
a superpotential of $W = mX\phi$ is introduced, where $\phi$ denotes the inflaton and 
$X$ is another superfield.
Also in this case $X$ is settled to zero during inflation and the constant term in the superpotential
does not have important effects.


\section*{Acknowledgment}

This work was supported by the Grant-in-Aid for Scientific Research 
from the Ministry of Education, Science, Sports, and Culture (MEXT) (No. 21111006) [KN],
 the Grant-in-Aid for Scientific Research on Innovative Areas (No. 21111006) [FT],  Scientific Research (A)
(No. 22244030 [FT] and 22244021 [TTY] ), and JSPS Grant-in-Aid for Young Scientists (B) (No. 21740160) [FT]. 
  This work was also supported by World Premier
  International Center Initiative (WPI Program), MEXT, Japan.


{}

\end{document}